\documentclass[aps,prl,twocolumn]{revtex4-1}

\usepackage{graphicx}
\usepackage{amssymb}
\usepackage{amsmath}
\usepackage{cleveref}
\usepackage{bbold}

\crefname{equation}{Eq.}{Eqs.}
\crefname{figure}{Fig.}{Figs.}

\renewcommand{\eqref}[1]{Eq. (\ref{#1})}
\newcommand{\figref}[1]{Fig. \ref{#1}}
\newcommand{\bms}[1]{\boldsymbol{#1}}

\newcommand{\vphi}{\varphi}

\newcommand{\Ui}{U_{\infty}}
\newcommand{\Ut}{U_{\infty}}
\newcommand{\Ti}{T_{\infty}}
\newcommand{\bn}{\boldsymbol{n}}
\newcommand{\pef}{{\rm Pe}_\vphi}
\newcommand{\pet}{{\rm Pe}_T}
\newcommand{\ca}{{\rm Ca}}
\newcommand{\ch}{{\rm C_h}}
\newcommand{\lew}{{\rm Le}}
\newcommand{\tv}{\boldsymbol{v}}
\newcommand{\tT}{\Theta}
\newcommand{\tnabla}{\nabla}

\newcommand{\tTc}{\tT_c}
\newcommand{\tTw}{\tT_{\rm cov}}

\begin{document}

\title{Self-propulsion mechanism of active Janus particles in near-critical 
binary mixtures}

\author{Sela Samin}
\email{S.Samin@uu.nl}
\author{Ren\'{e} van Roij}

\affiliation{Institute for Theoretical Physics, 
Center for Extreme Matter and Emergent Phenomena, 
Utrecht University, Leuvenlaan 4, 3584 CE Utrecht, The Netherlands}

\date{\today}


\begin{abstract}

Gold-capped Janus particles immersed in a near-critical binary mixture can be 
propelled using illumination. We employ a non-isothermal diffuse interface 
approach to investigate the self-propulsion mechanism of a single colloid. We 
attribute the motion to body forces at the edges of a micronsized droplet that 
nucleates around the particle. Thus, the often-used concept of a 
surface velocity cannot account for the self-propulsion. The particle's 
swimming velocity is related to the droplet shape and size, which is determined 
by a so-called critical isotherm. Two distinct swimming regimes exist, 
depending on whether the droplet partially or completely covers the particle. 
Interestingly, the dependence of the swimming velocity on temperature is 
non-monotonic in both regimes. 
\end{abstract}
\maketitle

The study of self-propelling synthetic colloids is an area of 
intense active research \cite{ebbens2010,aranson2013}. The 
out-of-equilibrium directed motion of these
colloidal microswimmers is maintained by a constant 
energy input which originates from their own activity. The directed swimming, 
coupled to the particle's rotational diffusion, leads to a significant increase 
in the effective diffusion coefficient 
\cite{howse2007,jiang2010,volpe2011} and to complex collective 
behavior, 
such as dynamical phase-separation \cite{buttinoni2013,stenhammar2013,speck2014} 
and clustering \cite{theurkauff2012,palacci2013,pohl2014}. Optimization of the 
microswimmers design is essential for realizing applications such as targeted 
cargo and drug delivery, parallel assembly and scavenging 
of contaminants \cite{ebbens2010,popescu2011,baraban2012}.

The design of synthetic swimmers requires an understanding of 
the underlying mechanisms for the self-propulsion, 
\textit{e.g.} self-diffusiophoresis 
\cite{howse2007,kapral2013,golestanian2005,cordova2008,brady2010,graaf2015}, 
self-induced 
electrophoretic flow, \cite{paxton2004,moran2011} and self-thermophoresis 
\cite{jiang2010,bickel2013}. In many realizations, the particle motion is 
attributed to a microscopically thin 
boundary layer adjacent to the solid-fluid interface, which interacts with a 
self-generated field, such as electrical potential, solute concentration and 
temperature. Body forces within this layer give rise to an apparent slip 
velocity at the surface \cite{anderson1989} while the fluid 
outside the interfacial layer is 
considered force-free. Thus, the particle motion is completely 
determined by the slip velocity distribution on its surface 
\cite{brady2010,bickel2014}. However, this simple picture breaks down when the 
self-generated field extends to a region with a size similar to that of the 
particle. In this Letter we explore such a scenario of 
self-diffusiophoresis due to a local solvent demixing, leading to a complex 
swimming behavior arising from the coupling of the self-generated 
chemical potential gradients and the fluid motion. 

\begin{figure}[!b]
\includegraphics[width=2.75in,clip]{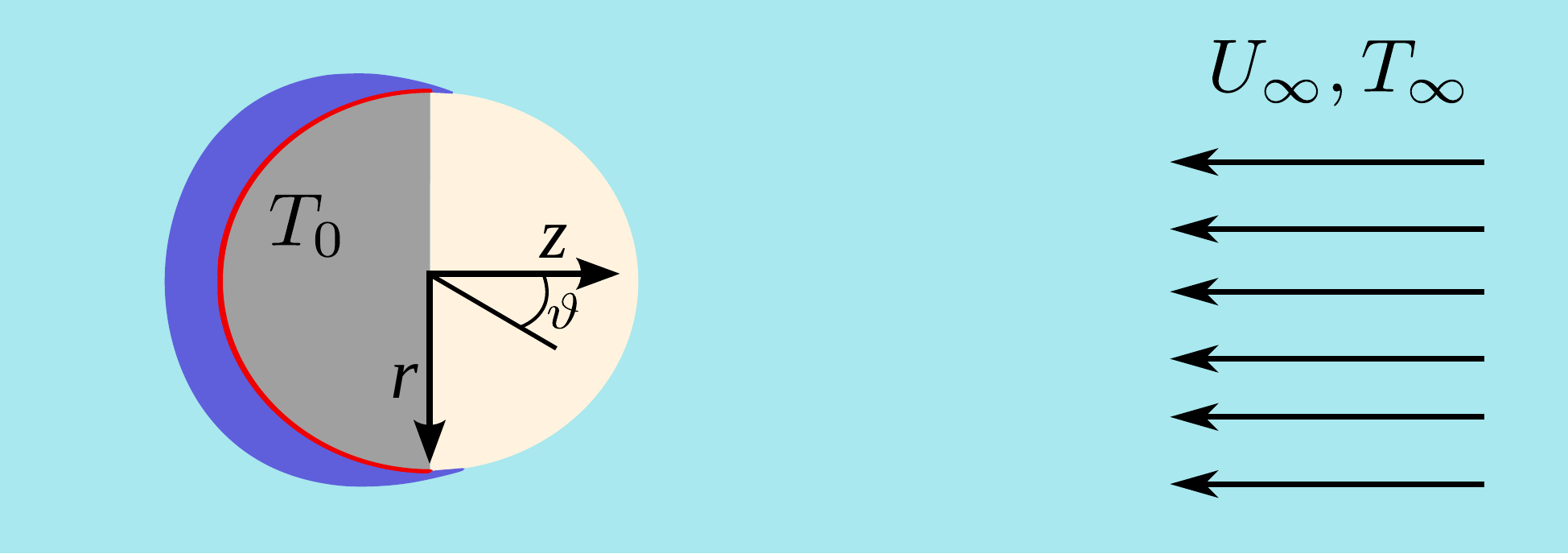}
\caption{Schematic illustration of a Janus particle immersed in 
a near-critical binary mixture at a temperature $\Ti<T_c$. Illumination 
results in $T_0>T_c$ at the gold-capped hemisphere (red line). At steady state, 
a demixed droplet is nucleated around the translating colloid, with a 
fluid velocity $\Ui$ at infinity in the frame-of-reference 
co-moving with the particle.}
\label{fig_sys}
\end{figure}

We focus on a recently realized new class 
of swimmer consisting of Janus colloids 
immersed in a near-critical binary mixture. 
Local heating of the colloid surface and the ensuing solvent demixing propels 
these particles, which exhibit fascinating individual and collective 
behaviour \cite{buttinoni2013,buttinoni2012,hagen2014,kuemmel2013,volpe2011}. A 
similar system was studied by Araki 
and Fukai \cite{araki2015} but in their simulations heating is periodically 
applied to the whole mixture. In this work we study the 
self-propulsion mechanism of a locally heated Janus swimmer (illustrated in 
\cref{fig_sys}) and show that it 
is 
fundamentally different from motion driven by interfacial velocities. Rather, 
we find that the particle motion is 
linked to the flow at the edges of a stationary demixed droplet, also
far from the particle surface, and that it depends strongly on 
the solvent and particle properties. These advective effects are ignored in 
W\"urger's recent study of a similar system \cite{wurger2015}.

We consider a micronsized spherical colloid with radius $R$ immersed in an 
unbounded homogeneous near-critical binary mixture. Half of the particle is 
gold-capped 
and continuously heated by irradiation. The gold layer thickness is
typically of the order of 10nm, much smaller than the colloid radius, and 
its thermal conductivity is much 
larger than that of the liquid and colloid. Thus, the cap forms an isotherm 
\cite{bickel2013} with a temperature 
$T_0>\Ti$, where $\Ti$ 
is the fluid's ambient temperature far from the colloid. 
Local demixing of the fluid occurs 
if the temperature around the colloid increases above the (lower) critical 
temperature $T_c$ of the mixture, of which both components are assumed to have 
a molecular volume $a^3$ with $a\sim3$\AA{}. Consequently, temperature 
gradients in the 
demixed phase 
lead to gradients of the mixture chemical 
potential which give rise to a local body-force. From azimuthal symmetry, no 
net torque 
will act on the colloid but since the temperature field is 
not spherically symmetric, the demixed droplet exerts a net force on the 
colloid in the axial direction. Hence, we set a cylindrical coordinate system 
in a 
frame-of-reference co-moving with a colloid placed at its origin. Our aim is to 
find the axial velocity of the fluid far away 
from the colloid: $\bms{\Ut}=-\Ui\hat{\bms{z}}$ such that the colloid is 
force-free. The main result of this paper is that the self-propulsion 
is a result of forces perpendicular to the colloid surface, and cannot be 
attributed to an effective surface velocity parallel to it, as is common in 
many 
scenarios \cite{bickel2013,kapral2013,brady2010,anderson1989}.

We investigate the steady-state during the ballistic motion of the 
colloid at a time scale much shorter than its rotational diffusion 
time \cite{buttinoni2012}. To this end we employ a well-established diffuse 
interface based approach \cite{anderson1998}, that couples the mixture order 
parameter $\vphi\in [-1/2,1/2]$, the mixture chemical potential $k_BT\mu$, the 
fluid 
velocity $\Ui\tv$, and the scaled temperature $\tT=(T-\Ti)/(T_0-\Ti)$ by the 
dimensionless equations:
\begin{align}
\label{eq:ch_dl}
 \partial\vphi / \partial t &=- \tnabla \cdot \left(\pef \vphi 
\tv-\tnabla \mu\right)~,\\
\label{eq:mu_dl}
\mu&= -\epsilon \nabla^2\vphi+f'(\vphi)~,\\
\label{eq:nc_dl}
\tnabla \cdot \tv &=0~,
\\
\label{eq:ns_dl}
\tnabla\cdot \tau &= 
\nabla p+ \ch^{-1}\ca^{-1}\vphi \tnabla \mu~, \\
 \lew^{-1} \partial\tT / \partial t &=- \tnabla \cdot 
\left( \pet \tT \tv -\tnabla \tT\right) 
~.
\label{eq:heat_dl}
\end{align}
Here, all lengths are scaled by $R$ and time is scaled by $R^2/D$, where $D$ is 
the mixture inter-diffusion constant. 

\cref{eq:ch_dl} is the convective
Cahn-Hilliard 
equation governing the composition dynamics, where the 
relative magnitude of the composition
advective flux, $\vphi \bms{v}$, and diffusive flux, $D \nabla \mu$, is 
measured by the 
the composition P\'{e}clet number, $\pef=\Ui R/D$. The chemical 
potential \cref{eq:mu_dl} is obtained via the bulk free energy 
$a^3f=(\chi-2)\vphi^2+4\vphi^4/3$, where $\chi \sim 1/T$ is the Flory 
interaction 
parameter, supplemented by the square-gradient term which accounts for 
interfacial tension \cite{supp}. The interfacial width is characterized by 
$\epsilon=\chi\ch^2$, where $\ch=a/R$ is the Cahn number. 
\cref{eq:ns_dl,eq:nc_dl} are 
the Stokes equations governing the fluid flow. Here, $p$ and 
$\tau=\nabla\bms{v}+\nabla\bms{v}^T$ 
are the dimensionless fluid pressure and viscous stress tensor, respectively, 
scaled 
by $\eta_f\Ui/R$, where $\eta_f$ is the 
fluid viscosity. The last term in 
\eqref{eq:ns_dl} is the body force due
to gradients in the chemical potential, it is proportional to the inverse 
capillary number, $\ca=a^2\eta_f\Ui/k_BT$, which measures the 
relative magnitude of viscous and surface tension forces.
\cref{eq:heat_dl} is 
the energy equation for the fluid. Here, 
$\pet=\Ui R/\alpha$ is 
the thermal P\'{e}clet number, where $\alpha_f=k_f/(\rho_f C_f)$ is the 
fluid thermal 
diffusivity, and $\rho_f$, $C_f$ and $k_f$ are the fluid density, heat capacity 
and thermal conductivity, respectively. The Lewis number, 
$\lew=\alpha_f/D$, is the ratio of thermal to mass diffusivity.

The force, $\bms{{\rm F}}$, exerted on the colloid by the fluid is estimated by 
applying 
the 
divergence theorem to \eqref{eq:ns_dl}:
\begin{align}
\bms{{\rm F}}=2\pi \int_{-1}^{1}{\rm d}c~c \left[p\mathbb{1}+\Pi-\tau 
\right]\cdot\bn~,
\label{eq:force}
\end{align}
where $c=\cos\vartheta$, $\vartheta$ is the polar angle and $\bn$ is 
the outward 
unit 
vector normal to the surface. By recasting $\vphi \nabla 
\mu=\nabla\cdot\Pi$ one obtains \cite{anderson1998} the Korteweg 
stress tensor,
% \begin{align}
$\Pi =\left[ 
\vphi f'(\varphi) -f-\tfrac{1}{2}
\epsilon|\nabla\vphi|^2-\epsilon\vphi\nabla^2\vphi\right] 
\mathbb{1}+\epsilon\nabla\vphi\nabla\vphi.$
% \label{eq:st}
% \end{align}

\begin{figure}[!t]
\includegraphics[width=3.5in,clip]{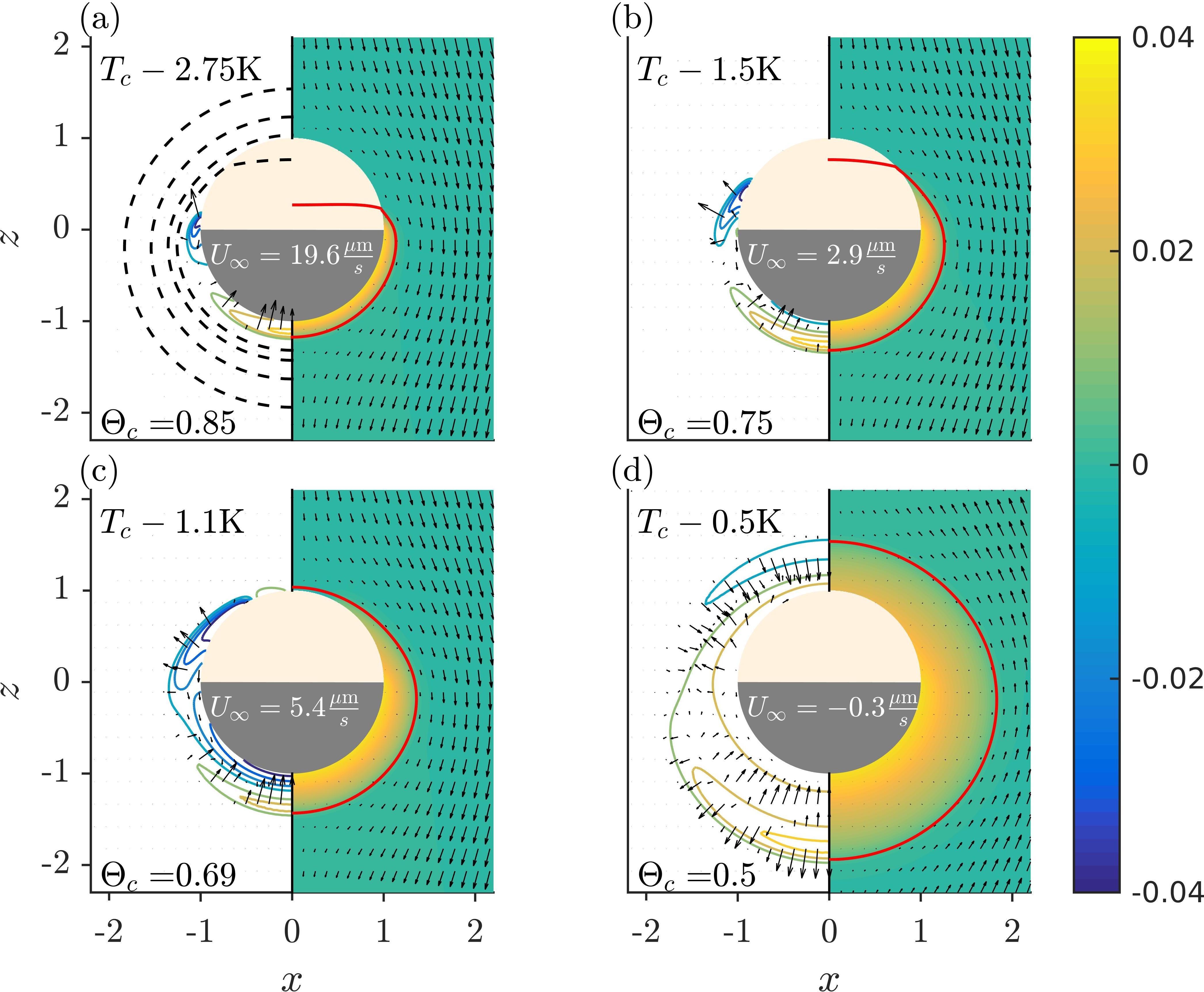}
\caption{Steady state composition $\vphi$ (right) and contours of the 
scaled pressure $p/200$ (left) around a heated Janus particle in the $xOz$ 
plane. The temperature of the heated cap is $T_0=T_c+0.5$K. The bulk mixture 
temperature approaches $T_c$ from (a) to (d) with $T_c-\Ti$ 
equal to (a) 2.75K, (b) 1.5K, (c) 1.1K and (d) 0.5K. The red curve is the 
critical contour $\tTc$. Arrows are vectors of the velocity 
(right) and body force [\eqref{eq:ns_dl}] (left). In (a) the 
dashed lines are the contours 
$\tT=0.75,0.69,0.6,0.5$. Here, $D=4\times 10^{-11}$ m$^2$/s, 
$\theta_1=\pi/4$ and $\theta_2=\pi/2$. For the physical properties of the 
2,6-lutidine-water mixture we take $a=3.4$\AA{}, $\rho_f=987$ kg/m$^3$, 
$\eta_f=2$ mPa s, 
$C_f=4.3$ kJ/(kg K) and $k_f=0.39$ W/(m K) 
\cite{grattoni1993,pittois2004,mirzaev2006}. For the silica particle we used 
$k_s=1.38$ W/(m K).}
\label{fig_ss1}
\end{figure}

The physical 
properties are taken to best mimic the 
experimental setup in Ref. \citep{buttinoni2012}. Therefore, we use properties 
of silica glass for the colloid and those of a critical water--2,6-lutidine for 
the fluid. Since $T_0 - \Ti$ 
is only of the order of 1K we can assume that all material properties are
constant \cite{supp}, see the caption of \figref{fig_ss1} for the
values used in the calculation. The assumption that $D$ is constant is a 
somewhat crude approximation, since it 
vanishes as a power law close to $T_c$ \cite{kawasaki1970}.
In the temperature window we examine $D$ is of the order of 
$10^{-11}-10^{-12}$m$^2$/s \cite{gaulari1972}. We therefore use an 
``effective'' 
value of $D$ as a parameter and examine its influence on the swimming. In 
addition, the exact value of the contact angle 
of each hemisphere in water--2,6-lutidine is unknown. We therefore use an 
indicative contact angle of $\theta_1=\pi/4$ for the hydrophilic gold cap 
throughout this work such that $\vphi>0$ corresponds to a 
water-rich phase. We will explore the influence of $\theta_2$ at the other 
hemisphere on 
the swimming.

Taking $R=0.5\mu$m, $D=4\times 10^{-11}$ m$^2$/s and a large $\Ui=20 
\mu$m/s 
leads to 
$\pef \approx 0.2$, $\pet \approx 10^{-4}$, $\lew \approx 10^3$, $\ch\approx 
10^{-4}$ and $\ca 
\approx 
10^{-5}$. Even for this extremal $\Ui$, $\pet \ll 1$ and we can 
safely neglect the heat 
advection in \cref{eq:heat_dl}. In addition, because heat 
diffuses much faster than mass, $\lew \gg1 $, the temperature adjusts almost
instantly to a composition perturbation and we may neglect also the time 
dependence in \cref{eq:heat_dl}, which leads to the heat 
equation $\nabla^2\tT=0$. The same argument holds for the solid where we 
also solve the Laplace equation. Hence, the temperature distribution 
\emph{only} 
depends on the thermal conductivity contrast of the solid and fluid, 
$k_s/k_f$, where $k_s$ is the colloid thermal conductivity. Therefore, we solve 
the heat equation once, and use 
the resulting temperature distribution as an input for the relaxation of the 
velocity and composition. 
This relaxation is dominated by the body force $\vphi\nabla\mu$ in 
\eqref{eq:ns_dl} since $\ca \ll 1$.

The colloid is placed at the 
origin of a large cylindrical domain of height $z\in[-L,L]$ and radius 
$r\in[0,L]$ with $L=500$ \cite{supp}. We use symmetry boundary 
conditions (BCs) at $r=0$. 
At the other edges of the domain we impose, using the appropriate BCs, a mixture
of critical composition at a temperature $\Ti$ ($\tT=0$) with a velocity 
$\bms{\Ut}$ 
\cite{supp}.
On both colloid hemispheres we impose a no-slip BC for the fluid: 
$\bms{v}_i=0$, where 
$i=1,2$ denotes the capped and uncapped hemisphere, respectively. At the capped 
hemisphere we set $\tT_1=1$ while for the uncapped hemisphere we 
a have continuity of the heat flux, $\bn \cdot k_f\nabla \tT_{out} =\bn \cdot 
k_s \nabla \tT_{in}$.
The first BC for the composition at the 
colloid surface is no flux: $\bn \cdot \nabla \mu=0$.
The colloid has two chemically distinct 
solid-liquid interfaces, for which we assume an excess surface 
free energy 
$F_w$ of the form $a^2F_w/k_BT=\sum_i\int \gamma_i\varphi dA_i$, where 
$\gamma_i$ 
measures the difference between 
the microscopic short-range interaction of the two solvent components and the 
solid. The 
wetting angles $\theta_i$ are then imposed using: $\bn \cdot \nabla \vphi = 
-\tan\left(\pi/2-\theta_i\right)\left| 
(\mathbb{1}-\bn\bn)\cdot\nabla \vphi \right|$,
where in this so-called geometric formulation of the wetting BC
$\gamma_i=\cos\theta_i/\sqrt{2}$ \cite{ding2007}. This BC has proved 
useful in 
simulations of moving contact lines where it is known to result in an effective 
slip through the diffusive
fluxes between the phases \cite{chen2000,jacqmin2000,ding2007}. Thus, fluid 
motion due to the interaction with the surface is actually resolved even 
though the no-slip BC is imposed.

Modern day computational resources coupled 
to specialized mesh generation allowed us to resolve the fluid spatial 
distribution over 6 orders of magnitude, from the microscopic scale set by the 
interfacial width, $O(a)$, up to the mesoscopic scale set by the domain 
size $L$ \cite{supp}.

\noindent \textit{Steady State.} The resulting composition of the mixture 
around 
a 
force-free 
swimmer ($\bms{{\rm F}}=0$) for several $\Ti$ and 
fixed $T_0=T_c+0.5$K is shown in \cref{fig_ss1} (a)-(d). The solid red line in 
each panel is the contour of the reduced critical 
temperature, $\tTc=(T_c-\Ti)/(T_0-\Ti)$, which can account for many features 
of the swimming. Demixing 
only
occurs inside the region bounded by the $\tTc$ isotherm, where $T>T_c$. In 
\cref{fig_ss1} (a) 
we also show several other contours of $\tT<\tTc$ (dashed lines). Demixing 
within 
these contours will occur for fixed
$T_0>T_c$ and increasing $\Ti$ [see \cref{fig_ss1} (b)-(d)] or for fixed 
$\Ti$ and increasing $T_0$. In both cases $\tTc$ decreases and thus the droplet 
grows.

\cref{fig_ss1} also reveals that a single water-rich droplet is nucleated at 
the particle surface. Within the droplet 
the composition is inhomogeneous; $\vphi$ is maximal at the 
surface and decays 
rather smoothly to the bulk value $\vphi=0$ because 
of the temperature gradients and the proximity to $T_c$. The 
demixed region is clearly distinct from the bulk phase, as can be seen from the 
velocity vectors. Inside 
the demixed region the velocity is very small implying the droplet effectively 
moves 
together with the particle. Strikingly, we find that \emph{no} significant slip 
occurs at 
the 
surface, in contrast to the results in Ref. \cite{wurger2015}. The 
fluid weakly 
circulates inside the droplet \cite{supp} and the overall flow pattern is 
similar to that of Stokes flow past a viscous droplet \cite{pozrikidis2011}.

We distinguish between 
two droplet shapes: (i) when $\tTc<\tTw$ the droplet partially 
covers the particle as in \cref{fig_ss1} (a)-(c) and (ii) for $\tTc>\tTw$ 
complete covering occurs as in \cref{fig_ss1} (d). The covering temperature 
$\tTw$ is closely related to the uncapped pole isotherm $\tT_p=\tTc$, for which 
the demixed region should first encompass the particle. For the solid-fluid 
heat 
conductivity contrast $k_s/k_f\approx3.5$ 
that we use, $\tT_p \approx0.7$. However, a thin 
demixed region at the pole is energetically costly, and in fact our numerical 
solution gives $\tTw\approx0.66$ somewhat smaller than $\tT_p$ and a 
discontinuity in 
the pole composition at $\tTw$. In comparison, 
for $k_s/k_f=1$ the lower heat diffusivity in the 
solid leads to $\tT_p=0.5$ \cite{bickel2013}. Hence, the 
droplet shape and therefore the swimming behavior are both quite sensitive to 
the conductivity contrast, which we thus identify as an interesting engineering 
parameter.

To maintain the steady-state shape of the droplet as dictated by the contour 
$\tTc$, the composition diffusive flux $\propto \nabla\mu$ balances the 
convective flux according to 
\eqref{eq:ch_dl}. Therefore, the 
body force $\propto \vphi\nabla\mu$ is primarily large at the droplet diffuse 
boundary where advection becomes significant, see the vector maps 
in 
\figref{fig_ss1} (a)-(c). The body force may be large 
within the droplet when $\Ui$ is small and the droplet internal dynamics 
also becomes significant, see \figref{fig_ss1} (d). Notice that the 
pressure contours in \figref{fig_ss1} 
are 
approximately perpendicular to $\vphi\nabla\mu$ because within the droplet 
$\nabla\cdot\tau$ is small, and thus $\nabla 
p\propto\vphi\nabla\mu$ to first-order at steady-state. Moreover, at the front 
of the droplet (with respect to the particle direction of motion) the diffusive 
flux must balance an advective flux toward the droplet whereas at the rear 
of the droplet the advective flux carries the mixture away from the droplet. 
Thus, also the body force at the droplet edge acts in opposite directions 
relative to 
the fluid flow, resulting in the pressure distributions of \cref{fig_ss1} 
(a)-(c) exhibiting two regions: (i) a $p>0$ region at the droplet rear and (ii) 
a region of $p<0$ near the three 
phase contact line.

The anisotropy of the droplet shape produces an anisotropy 
also of the pressure and body force within the droplet, which is responsible 
for the particle motion. Thus, the self-propulsion is a result of forces 
perpendicular to the surface, which are affected by the flow far from the 
surface, and cannot be mapped to an effective surface velocity. In 
\cref{fig_ss1} (a)-(c), the resulting force-free motion is in the direction of 
the uncapped hemisphere whereas the completely covered colloid in 
\cref{fig_ss1} (d) moves with 
the capped hemisphere on the front.

\begin{figure}[!t]
\includegraphics[width=2.75in,clip]{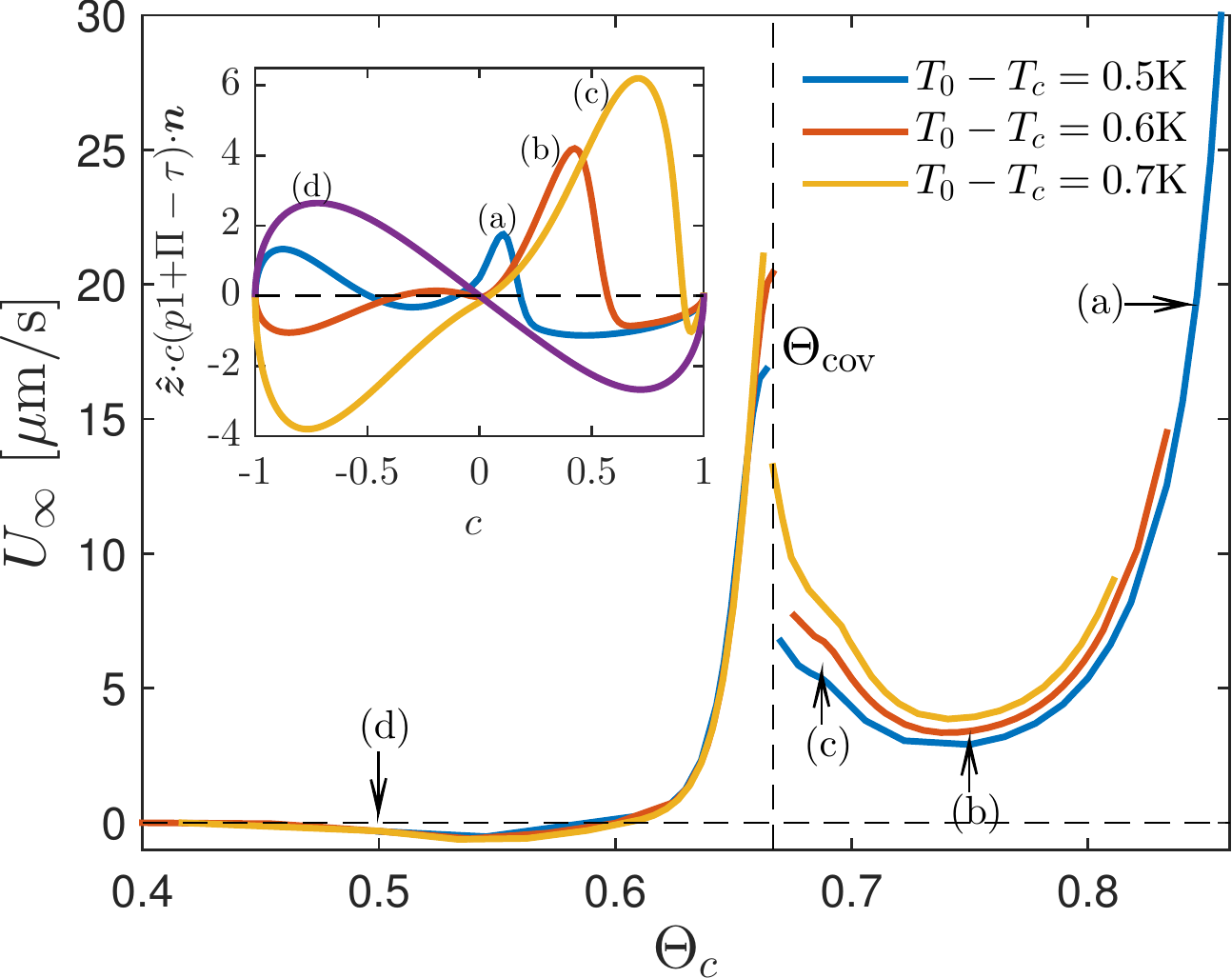}
\caption{Swimming velocity vs. reduced critical 
temperature curves for three quenching temperatures $T_0-T_c$. The droplet 
completely 
covers the colloid for $\tTc< \tTw\approx0.66$ whereby the 
different curves collapse onto one another. For $\tTc>\tTw$ the 
curves are distinct with the swimming velocity increasing with quenching. 
The velocity of the swimmers in \cref{fig_ss1} in indicated with arrows and the 
inset shows the corresponding force profiles along the colloid contour.}
\label{fig_vt_dt}
\end{figure}

\noindent \textit{Swimming Velocity.} \cref{fig_vt_dt} shows the swimming 
velocity $\Ui$ as a function of $\tTc$ 
for three values of the quenching $T_0-T_c$, which increases with the laser 
intensity in experiments. $\Ui$ 
strongly depends on
$\tTc$ and is of the order of $0.1-10\mu$m/s, in agreement with experiments. 
Two 
swimming regimes exist, for 
complete coverage at $\tTc<\tTw$ the swimming is independent of the quenching as
$\tTc$ completely determines the demixed state.
In contrast, for $\tTc>\tTw$ the curves are distinct 
with $\Ui$ increasing with quenching. This is because here $\tTc$ meets the 
particle surface at an angle different from the contact angle, which leads to a 
competition between the demixing and surface energies close to 
the three phase contact line, with the balance shifted in 
favor of the demixing as quenching increases. Experiments are performed at a 
single constant $\Ti$, showing that the swimming velocity increases with 
the laser intensity \cite{buttinoni2012}, but our calculations, which explore a 
large range of $\Ti$, indicate this is not always the case.

\cref{fig_vt_dt} also reveals that $\Ui$ jumps at $\tTw$ and is non-monotonic 
with $\tTc$ in both regimes. To 
understand this behaviour we plot in the inset of \figref{fig_vt_dt} the $z$ 
component of the force [cf. \cref{eq:force}] along 
the particle contour, $\bms{\hat{z}}\cdot c(p\mathbb{1}+\Pi-\tau)\cdot\bn$, as 
function of $c$, where $c<0$ ($c>0$) corresponds the 
(un)capped hemisphere. The labels in the inset and the arrow labels
in \cref{fig_vt_dt} correspond to the swimmers in the panels of \cref{fig_ss1}.
When $\tTc$ and $\Ui$ are large, as in \cref{fig_ss1} (a), the force profile 
has a positive maximum in each hemisphere. This is because the droplet is small 
and hence a large body force exists close to the surface at both hemispheres. 
When $\tTc$ decreases and the droplet grows, as in \cref{fig_ss1} (b)-(c), 
the positive maximum at the capped hemisphere becomes a negative minimum while 
the 
maximum at the uncapped hemisphere grows in magnitude and moves to a larger 
$c$. This is 
a result of the anisotropic shape of the droplet. While the
$\tTc$ contour is distanced from the capped hemisphere it remains close to the 
uncapped 
hemisphere and also covers a larger portion of it. Thus, the body force 
localized at the droplet edge becomes more significant near the three phase 
contact line rather than near the particle rear, thereby accounting for the 
minimum in $\Ui$. For $\tTc<\tTw$, $\Ui$ first jumps to a larger positive value 
but decreases with $\tTc$ and eventually becomes 
negative, since now the droplet edge at the front is also distanced from 
the particle. Finally, $\Ui$ vanishes when $\tTc$ is small and the large 
droplet becomes quasi-spherical ($\Ti \rightarrow T_c$).

The nucleation of a second, water-poor droplet, at the uncapped 
hemisphere is unfavored unless the hemisphere is very hydrophobic. 
The small volume of the demixed region around the uncapped hemisphere compared 
to the capped counterpart entails a relatively larger surface energy 
penalty. Nonetheless, we find that for $\theta_2\gtrapprox0.8\pi$ a water-poor 
droplet does nucleate at the uncapped hemisphere. This is accompanied by a 
reversal of the swimming direction \cite{supp}. Assuming that the 
uncapped silica is hydrophilic \cite{buttinoni2012,gallagher1992}, this 
result is consistent with the experimental observation that particles with a 
hydrophilic gold cap swim with their cap at the rear whereas for a 
hydrophobic cap it is at the front \cite{buttinoni2012}.

We also examined the dependence of $\Ui$ on the diffusion constant.
For a swimmer with the same parameters as in \cref{fig_ss1} (b) but with 
$D$ 10-fold larger or smaller, the resulting velocity is $111\mu$m/s and 
$0.36\mu$m/s, respectively. As expected, $\Ui$ increases with $D$ because 
the diffusive current is able to compensate for larger advection 
while maintaining the droplet shape. Further work is required to 
understand the more realistic scenario where $D$ may vary in 
space due to thermal gradients near $T_c$.

In conclusion, we have shown that a locally heated Janus particle 
in a near-critical binary mixture is propelled by the chemical potential 
gradients at the diffuse interface of a nucleated droplet, arising from the 
balance of diffusive and advective fluxes. Therefore, the self-propulsion 
cannot be 
described by an effective surface velocity. We hope that our results will 
stimulate further experiments to uncover the details of the swimming mechanisms 
of these intriguing particles and possibly explore other microswimmers propelled 
by non-local self-generated fields.

We acknowledge fruitful discussions with C. Bechinger and J. de Graaf and 
financial support of a Netherlands Organisation for Scientific
Research (NWO) VICI grant funded by the Dutch Ministry
of Education, Culture and Science (OCW). This work is part of the D-ITP 
consortium, a program of the Netherlands Organisation for Scientific
Research (NWO) funded by the Dutch Ministry of Education, Culture and Science 
(OCW).

\end{document}